# Affordability, cost and cost-effectiveness of universal anti-retroviral therapy for HIV

Brian G. Williams* and Eleanor Gouws†

* South African Centre for Epidemiological Modelling and Analysis, Stellenbosch, South Africa
† Joint United Nations Programme on HIV/AIDS (UNAIDS), Geneva, Switzerland

Correspondence to BrianGerardWilliams@gmail.com

**Abstract**

If people at risk of HIV infection are tested annually and started on treatment as soon as they are found to be HIV-positive it should be possible to reduce the case reproduction number for HIV to less than one, eliminate transmission and end the epidemic. If this is to be done it is essential to know if it would be affordable, and cost effective. Here we show that in all but eleven countries of the world it is affordable by those countries, that in these eleven countries it is affordable for the international community, and in all countries it is highly cost-effective.

## Introduction

The science behind *Treatment-as-Prevention* is clear.[1] Annual testing and immediate ART can stop transmission[2-7] and at the same time gives people infected with HIV the best prognosis.[8-16] Furthermore, a number of studies have shown that ART is cost-effective[17-19] even in poorly resourced countries. Here we use data on the number of people living with HIV and the gross domestic product (GDP), for all countries, to explore the affordability, cost and cost-effectiveness of putting all HIV-positive people onto ART.

## Methods

To estimate the affordability of ART we calculate, for each country in the world, an affordability index, AI, which is the cost of giving all HIV-positive people ART expressed as a percentage of GDP. If the AI is greater than 2% we regard universal ART as unaffordable, between 1% and 2% as marginal and less than 1% as affordable. To calculate the affordability if funding in resource limited countries is provided by the international community, we compare the cost of ART to current global spending on ART in lower and middle income countries. Since there is no universally agreed definition of affordability we also compare the cost of universal ART to the cost of spending on the military in each country. To estimate the cost-effectiveness we calculate the cost-effectiveness ratio, CER, as the cost of maintaining one person on ART as a proportion of the *per capita* GDP. If the cost per life year saved is less then three times *per capita* GDP we regard it is cost-effective, if it is less than the *per capita* GDP we regard it as highly cost-effective.[20]

We use data on the number of people living with HIV for each country in the world from UNAIDS.[21] We use data on the GDP from *Knoema*,[22] on the *per capita* GDP from *Wikipedia*[23] and on military spending from *The Guardian*.[24]

We assume that the cost of maintaining one person on ART for one year is US$500: half for drugs and half for monitoring, care and support.[25]

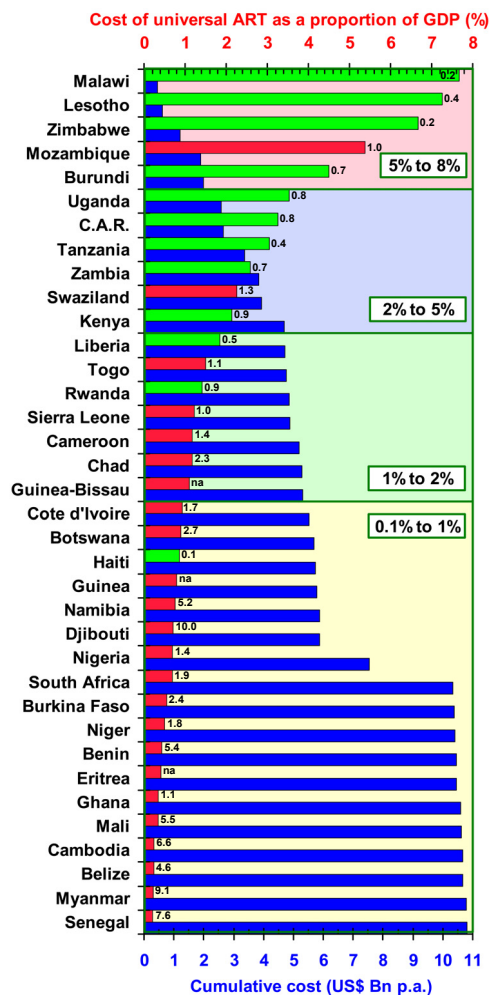

Figure 1. Red and green bars, upper axis: affordability index, AI: the cost of universal ART as a percentage of GDP. Blue bars, lower axis: cumulative cost of universal ART. Shaded areas: AI 5% to 8%, 2% to 5%, 1% to 2% and 0.1% to 1%. In all other countries the AI is less than 0.1%. Embedded numbers: military spending divided by the cost of universal ART. Red bars military spending greater than the cost of universal ART; green bars military spending less than the cost of universal ART.

## Results

In Figure 1 countries are ordered according to decreasing values of the AI. In Malawi, for example, the cost of maintaining all HIV-positive people on ART would amount to almost 8% of GDP which is unaffordable and the same holds for ten other countries for which the AI is greater than 2%: Lesotho, Zimbabwe, Mozambique, Burundi, Uganda, the Central African Republic, Tanzania, Zambia, Swaziland and Kenya. These countries will need international assistance to finance universal ART.

In seven countries, Liberia, Togo, Rwanda, Sierra Leone, Cameroon, Chad and Guinea-Bissau, the AI is between 1% and 2% of GDP and they will also need international assistance.

For the remaining 18 countries the AI is less than 1% and they should be able to finance ART out of their own budgets although support for technical assistance may still be needed. For all of the countries not shown in Figure 1 the AI is less than 0.1% and these countries should be able to maintain all HIV-positive people on ART.

In 2009 a total of US$ 16 billion was available for the AIDS response in lower and middle income countries[21] and the target is to increase this to US$ 22 billion.[26] The cumulative cost of universal ART is shown by the blue bars in Figure 1. For the first eleven countries for which international assistance is needed the total cost of universal ART is US$ 4.7 billion per year or 29% of 2009 global expenditure on ART in middle and lower income countries.[21] For all the countries shown in Figure 1 the total cost is US$ 10.8 billion per year, or 68% of 2009 global expenditure on ART in lower and middle income countries,[27] US$ 4.4 billion, or 41% of this, is needed for Nigeria and South Africa alone.

Only two countries, Zimbabwe and Liberia, have a CER close to 1 (data not shown) so that Universal ART is highly cost-effective in even these two countries, while in the other countries shown in Figure 1 the median CER is 0.33 (IQR: 0.21−0.49; Max: 1.10; Min 0.03).

In Figure 1 the ratio of military spending to the cost of universal ART is given by the embedded numbers; where this ratio is less than 1 the bars are green, where it is greater than 1 the bars are red. In almost all of the countries where the cost of universal ART is less then 2% of GDP, current spending more on their military is greater than the cost of universal ART. The two most important countries in this regard are South Africa and Nigeria which, as noted above, account for 40% of the global cost of universal ART. However, as shown in Figure 1, the cost of universal ART in both Nigeria and South Africa is about 0.7% of GDP; in Nigeria this amounts to only 70% of current military spending and in South Africa to only 50% of current military spending.

## Conclusions

In all but eleven countries universal ART should be affordable by the governments of those countries although support for technical assistance will still be needed. In the eleven countries in which universal ART is probably not affordable by their governments, the total cost of universal ART is only 29% of current global expenditure on HIV so that with international assistance universal ART is affordable in these countries as well. Even if the international community were to pay for ART in all those countries of the world where the AI is greater than 0.1% the total cost would only amount to 68% of current global expenditure on HIV and to 40% if we exclude Nigeria and South Africa. Since universal ART is highly cost-effective in every country in the world, based on the definition used by the World Health Organization,[20] financial considerations should not deter the world from working towards universal ART.

Finally, it is worth noting that the cost of ART in the United States is one hundred times the cost of ART in middle and lower income countries. Even allowing for this the AI in the United States is 0.2% making it affordable. The CER is 0.5 making it highly cost effective but the total cost is US$ 30 billion per year, one third more than the current spending on HIV in the United States of US$22 billion.[27]

Universal ART is affordable, cost-effective and within current global funding commitments. Since universal ART would also stop transmission[4] steps should be taken to start implementing universal ART as soon as possible.